\documentclass{ifacconf}

\makeatletter
\let\old@ssect\@ssect % Store how ifacconf defines \@ssect
\makeatother

\usepackage{natbib}								% required for bibliography (IFAC)
\usepackage[style=american]{csquotes}

\usepackage{booktabs}							% Better tables
\usepackage{tabularx}							% Tables

\usepackage{supertabular}
\usepackage{multirow}

\usepackage{nicefrac}							% Fractions
\usepackage{siunitx}							% Measurement Units
\usepackage{amssymb}
\usepackage{amsmath}
\usepackage{gensymb}
\usepackage{mathtools}							% For new math symbols such as coloneqq (Felix)
\usepackage{graphicx}							% Already loaded in the class "irsdiplom"
\usepackage{xcolor}
\usepackage{tikz}
\usepackage{pgfplots}
\usepackage{grffile}
\usepackage[american, nooldvoltagedirection]{circuitikz}				% for circuit diagrams (Bertus) for more info, see guide at (http://texdoc.net/texmf-dist/doc/latex/circuitikz/circuitikzmanual.pdf)
\usepackage[german,USenglish,UKenglish]{babel}
\usepackage[T1]{fontenc}
\usepackage{bm}									% bold and italic for symbols (Lukas)
\usepackage{dsfont}								% nicer set notations (Lukas)
\usepackage{lmodern}	%% Ersetzt das veraltete \package{caption} (Alte Warnung) aus irsdiplom.cls Zeile 1256 bis 1265

\usepackage{xspace}
\usepackage{enumerate}							% personalized enumeration (Felix)
\usepackage{cancel}								% cancel out parts of an equation
\usepackage[normalem]{ulem}						% \sout{} for strikethrough text, modifier keeps normal \emph{}
\usepackage{thmtools}							% for more powerful theorem controls and list of theorems
\usepackage[most]{tcolorbox}					% text colour boxes for use in theorems and definitions
\usepackage{enumitem}							% for resuming enumerated lists
\usepackage{hyphenat}							% To hyphenate already hyphenated words

\usepackage{xifthen}							% provides \isempty test

\usetikzlibrary{external}
\usetikzlibrary{shapes,arrows}
\usetikzlibrary{calc}
\usetikzlibrary{angles}
\usetikzlibrary{positioning}
\usetikzlibrary{fit}
\usetikzlibrary{backgrounds}
\usetikzlibrary{mindmap,trees}

\pgfplotsset{compat=newest}
\usetikzlibrary{plotmarks}
\usetikzlibrary{arrows.meta}
\usepgfplotslibrary{patchplots}

\usepackage{arydshln}					% for dashed horizontal lines (Bertus)
\usepackage{hyperref}					% Must be loaded last

\makeatletter
\def\@ssect#1#2#3#4#5#6{%
	\NR@gettitle{#6}% Insert key \nameref title grab
	\old@ssect{#1}{#2}{#3}{#4}{#5}{#6}% Restore ifacconf's \@ssect
}
\makeatother
	\definecolor{purpleDark}{RGB}{118, 4, 205}
	\definecolor{purpleLight}{RGB}{186, 102, 250}
	\definecolor{blueDark}{RGB}{52, 78, 243}
	\definecolor{blueLight}{RGB}{118, 135, 244}
	\definecolor{redDark}{RGB}{197, 34, 0}
	\definecolor{redLight}{RGB}{255, 91, 57}
	\definecolor{yellowDark}{RGB}{255, 183, 0}
	\definecolor{yellowLight}{RGB}{255, 204, 77}
	\definecolor{greenDark}{RGB}{0, 143, 53}
	\definecolor{greenLight}{RGB}{42, 189, 97}

	\colorlet{greenFaint}{green!10!white}
	\colorlet{redFaint}{red!10!white}

	\definecolor{redText}{RGB}{222, 2, 10}
	\definecolor{orangeText}{RGB}{245, 86, 0}
	\definecolor{greenText}{RGB}{20,125,50}
	\definecolor{blueText}{RGB}{0, 114, 190}%{rgb}{0.00000,0.44700,0.74100}
	\definecolor{purpleText}{RGB}{115, 38, 146}
	\definecolor{pinkText}{RGB}{255, 107, 250}

	\definecolor{MatlabGreen}{rgb}{0,0.6,0}
	\definecolor{MatlabGray}{rgb}{0.5,0.5,0.5}
	\definecolor{MatlabMauve}{rgb}{0.58,0,0.82}
	
	\colorlet{shadecolor}{gray!15}
	
	\colorlet{colorPhs}{blueText}						% Highlight PHS models in table
	
	\colorlet{colorIFeed}{blueText}						% Control overview: grid-feeding inverters
	\colorlet{colorIForm}{orangeText}					% Control overview: grid-forming inverters
	\colorlet{colorISupport}{greenText}					% Control overview: grid-supporting inverters
	\colorlet{colorIFeedForm}{purpleText}				% Control overview: combination (feeding and forming)
	\colorlet{colorIFormSupport}{purpleText}			% Control overview: combination (forming and supporting)
	\colorlet{colorIFeedFromSupport}{purpleText}		% Control overview: combination (all)

	\colorlet{colorSLyapDirect}{redText!90!white}
	\colorlet{colorSLyapLaSalle}{orangeText!80!white}
	\colorlet{colorSLyapEquation}{pinkText}
	\colorlet{colorSSynchronisation}{blueText}
	\colorlet{colorSSmallSignal}{purpleText}
	\colorlet{colorSDroopAssumed}{greenText}
	\colorlet{colorSOther}{black}
	
	\colorlet{colorDguComp}{greenDark}					% Colour of the dgu components
	\colorlet{colorLineComp}{blueDark}					% Colour of the line components
	\colorlet{colorLoadComp}{redDark}					% Colour of the load components
	\colorlet{colorLoadLineComp}{purpleDark}			% Colour of the load with line components added
	
	\colorlet{colorChange}{blueDark}
	\colorlet{colorCompensate}{redDark}

	\definecolor{colorSim1}{rgb}{0.00000,0.24000,0.72000}
	\definecolor{colorSim2}{rgb}{0.85000,0.32500,0.09800}
	\definecolor{colorSim3}{rgb}{0.92900,0.69400,0.12500}
	\definecolor{colorSim4}{rgb}{0.49400,0.18400,0.55600}
	\definecolor{colorSim5}{rgb}{0.00000,0.61000,0.52870}
	\colorlet{colorPhasorLoadModel}{greenDark}
	\definecolor{colorPhasorHelpVars}{rgb}{0.85000,0.32500,0.09800} % red
	\definecolor{colorPhasorDesiredVars}{rgb}{0.00000,0.44700,0.74100} % blue
	\definecolor{colorPhasorMeasuredVars}{rgb}{0.49400,0.18400,0.55600} % purple

	\addto\extrasUKenglish{%
	}

	\addto\extrasUSenglish{%
	}

	\newcommand{\citeColorTxt}[3]
		{{\color{#1}\hypersetup{citecolor=#1}%
			\if\relax #2\relax% If statement for checking empty parameter: https://tex.stackexchange.com/questions/53068/how-to-check-if-a-macro-value-is-empty-or-will-not-create-text-with-plain-tex-co
				\cite{#3}% if #2 is empty
			\else%
				\cite[#2]{#3}% if #2 is not empty
			\fi%
			\hypersetup{citecolor=black}}}
		
	\makeatletter
	\newcommand\autorefMulti[1]{\@first@ref#1,@}
	\def\@throw@dot#1.#2@{#1}% discard everything after the dot
	\def\@set@refname#1{%    % set \@refname to autoefname+s using \getrefbykeydefault
		\edef\@tmp{\getrefbykeydefault{#1}{anchor}{}}%
		\xdef\@tmp{\expandafter\@throw@dot\@tmp.@}%
		\ltx@IfUndefined{\@tmp autorefnameplural}%
			{\def\@refname{\@nameuse{\@tmp autorefname}s}}%
			{\def\@refname{\@nameuse{\@tmp autorefnameplural}}}%
	}
	\def\@first@ref#1,#2{%
		\ifx#2@\autoref{#1}\let\@nextref\@gobble% only one ref, revert to normal \autoref
		\else%
			\@set@refname{#1}%  set \@refname to autoref name
			\@refname~\ref{#1}% add autoefname and first reference
			\let\@nextref\@next@ref% push processing to \@next@ref
		\fi%
		\@nextref#2%
	}
	\def\@next@ref#1,#2{%
		\ifx#2@ and~\ref{#1}\let\@nextref\@gobble% at end: print and+\ref and stop
		\else, \ref{#1}% print  ,+\ref and continue
		\fi%
		\@nextref#2%
	}
	\makeatother

	\renewcommand{\dq}{\emph{dq}\xspace}				% dq frame
	\newcommand{\Matlab}{{\rm \sc Matlab}\xspace}			% Matlab
	\newcommand{\Simulink}{{\rm \sc Simulink}\xspace}		% Simulink
	\DeclareSIUnit{\pu}{pu}							% Per Unit
	\DeclareSIUnit{\VAR}{\volt\ampere{}R}			% VAR

	\newcommand{\addWithPreComma}[1]{%
		\if\relax #1\relax% If statement for checking empty parameter
		\else%
		,#1% if #1 is not empty
		\fi%
	}
	\newcommand{\addWithPostComma}[1]{%
		\if\relax #1\relax% If statement for checking empty parameter
		\else%
		#1,% if #1 is not empty
		\fi%
	}
	\newcommand{\addInParentheses}[1]{%
	\if\relax #1\relax% If statement for checking empty parameter
	\else%
		(#1)% if #1 is not empty
	\fi%
	}

	\usepackage{listings}  
\theoremstyle{plain}
\newtheorem{lemma}{Lemma}
\newtheorem{assumption}{Assumption}
\newtheorem{proposition}{Proposition}
\theoremstyle{definition}
\newtheorem{remark}{Remark}

	\renewcommand{\vec}[1]{\bm{#1}}				% Vector underline
	\renewcommand{\matrix}[1]{\bm{#1}}		% Matrix underline			
	
	\newcommand{\Reals}{\ensuremath{\mathbb{R}}}		% Reals
	\newcommand{\imag}{\ensuremath{\text{i}}}			% Imaginary number
	
	\newcommand{\Transpose}{\ensuremath{\textsf{T}}}	% Transpose sign
	\newcommand\qedsymbol{$\blacksquare$}				% QED symbol
	
	\newcommand{\dPartial}[2]{\dfrac{\partial #1}{\partial #2}}					% Partial derivative
	\newlength\mytemplena
	\newlength\mytemplenb
	\DeclareDocumentCommand\myalignalign{sm}
	{
		\settowidth{\mytemplena}{$\displaystyle #2$}%
		\setlength\mytemplenb{\widthof{$\displaystyle=$}/2}%
		\hskip-\mytemplena%
		\hskip\IfBooleanTF#1{-\mytemplenb}{+\mytemplenb}%
	}

	\newcommand{\denoteHamil}{H}
	\newcommand{\denoteCurrent}{I}
	\newcommand{\denoteVoltage}{V}
	\newcommand{\denoteError}{\epsilon}
	\newcommand{\denoteStateError}{\xi}

	\newcommand{\denotePolyLoadCoeff}{a}

	\newcommand{\sD}{\text{d}}
	\newcommand{\sQ}{\text{q}}
	\newcommand{\sDQ}{\text{dq}}
	
	\newcommand{\sLoad}{\text{L}}
	
	\newcommand{\sMicrogrid}{\text{M}}

	\newcommand{\x}[1][]{\vec{x}_{#1}}
	\newcommand{\y}[1][]{\vec{y}_{#1}}
	\newcommand{\vu}[1][]{\vec{u}_{#1}}
	\newcommand{\vd}[1][]{\vec{d}_{#1}}
	\newcommand{\vdRef}[1][]{\vec{d}_{#1}^*}
	\newcommand{\z}[1][]{\vec{z}_{#1}}
	\newcommand{\zRef}[1][]{\vec{z}_{#1}^*}
	\newcommand{\f}[2]{\vec{f}_{#1}\addInParentheses{#2}}
	
	\newcommand{\eig}[1][]{\lambda_{#1}}

	\newcommand{\dfdu}[1][]{\nabla\f{}{{\vu[#1]}}}

	\newcommand{\vecError}[1]{\vec{\denoteError}_{#1}}
	
	\newcommand{\dVecError}[1]{\dot{\vec{\denoteError}}_{#1}}
	\newcommand{\stateError}[1][]{\vec{\denoteStateError}_{#1}}
	\newcommand{\stateErrorL}[1][]{\stateError[\sLoad\addWithPreComma{#1}]}

	\newcommand{\xL}[1][]{\x[\sLoad\addWithPreComma{#1}]}

	\newcommand{\xRef}[1][]{\x[#1]^*}
	\newcommand{\xLRef}[1][]{\xRef[\sLoad\addWithPreComma{#1}]}
	\newcommand{\xMRef}[1][]{\xRef[\sMicrogrid\addWithPreComma{#1}]}
	\newcommand{\yRef}[1][]{\y[#1]^*}
	\newcommand{\vuRef}[1][]{\vu[#1]^*}
	\newcommand{\J}{\matrix{J}}

	\newcommand{\K}{\matrix{K}}

	\newcommand{\A}{\matrix{A}} % A square matrix used in the fundamentals section
	\newcommand{\JL}[1][]{\J_{\sLoad\addWithPreComma{#1}}}

	\newcommand{\KL}[1][]{\K_{\sLoad\addWithPreComma{#1}}}

	\newcommand{\setU}{\mathcal{U}}
	\newcommand{\setY}{\mathcal{Y}}

	\newcommand{\Hamil}[3]{\denoteHamil_{#1}^{#2}\addInParentheses{#3}}
	\newcommand{\HL}[3]{\Hamil{\sLoad\addWithPreComma{#1}}{#2}{#3}}
	\newcommand{\HLx}[1][]{\Hamil{\sLoad\addWithPreComma{#1}}{}{{\xL[#1]}}}
	\newcommand{\HLe}[1][]{\Hamil{\sLoad\addWithPreComma{#1}}{}{{\stateErrorL[#1]}}}
	\newcommand{\HLdot}[2]{\dot{\denoteHamil}_{\sLoad\addWithPreComma{#1}}\addInParentheses{#2}}
	
	\newcommand{\HLedot}[1][]{\HLdot{#1}{{\stateErrorL[#1]}}}

	\newcommand{\Id}[1][]{\denoteCurrent_{\sD\addWithPreComma{#1}}}
	\newcommand{\Iq}[1][]{\denoteCurrent_{\sQ\addWithPreComma{#1}}}
	\newcommand{\Idq}[1][]{\vec{\denoteCurrent}_{\sDQ\addWithPreComma{#1}}}

	\newcommand{\IPd}[1][]{\denoteCurrent_{\sD\addWithPreComma{#1}}}
	\newcommand{\IPq}[1][]{\denoteCurrent_{\sQ\addWithPreComma{#1}}}
	\newcommand{\IPdq}[1][]{\vec{\denoteCurrent}_{\sDQ\addWithPreComma{#1}}}
	
	\newcommand{\ILinedq}[1][]{\vec{\denoteCurrent}_{\sDQ\addWithPreComma{#1}}}
	
	\newcommand{\ILd}[1][]{\denoteCurrent_{\sLoad,\sD\addWithPreComma{#1}}}
	\newcommand{\ILq}[1][]{\denoteCurrent_{\sLoad,\sQ\addWithPreComma{#1}}}
	\newcommand{\ILdq}[1][]{\vec{\denoteCurrent}_{\sLoad,\sDQ\addWithPreComma{#1}}}
	
	\newcommand{\dILdV}[1][]{\dPartial{\ILdq[#1](\Vdq[#1])}{\Vdq[#1]}}
	
	\newcommand{\IdRef}[1][]{\Id[#1]^*}
	\newcommand{\IqRef}[1][]{\Iq[#1]^*}

	\newcommand{\Vd}[1][]{\denoteVoltage_{\sD\addWithPreComma{#1}}}
	\newcommand{\Vq}[1][]{\denoteVoltage_{\sQ\addWithPreComma{#1}}}
	\newcommand{\Vdq}[1][]{\vec{\denoteVoltage}_{\sDQ\addWithPreComma{#1}}}
	
	\newcommand{\Vddot}[1][]{\dot{\denoteVoltage}_{\sD\addWithPreComma{#1}}}
	\newcommand{\Vqdot}[1][]{\dot{\denoteVoltage}_{\sQ\addWithPreComma{#1}}}

	\newcommand{\VLd}[1][]{\denoteVoltage_{\sD\addWithPreComma{#1}}}
	\newcommand{\VLq}[1][]{\denoteVoltage_{\sQ\addWithPreComma{#1}}}
	\newcommand{\VLdq}[1][]{\vec{\denoteVoltage}_{\sDQ\addWithPreComma{#1}}}
	
	\newcommand{\VdRef}[1][]{\Vd[#1]^*}
	\newcommand{\VqRef}[1][]{\Vq[#1]^*}
	\newcommand{\VdqRef}[1][]{\Vdq[#1]^*}

	\newcommand{\Vamp}[1][]{\ensuremath{\denoteVoltage_{#1}}}

	\newcommand{\VampNominal}{\ensuremath{\denoteVoltage_{0}}}

	\newcommand{\freqRef}{\ensuremath{\omega_0}}
	
	\newcommand{\cR}[1][]{R_{#1}}
	
	\newcommand{\cL}[1][]{L_{#1}}
	\newcommand{\cC}[1][]{C_{#1}}

	\newcommand{\lAct}[1][]{P_\sLoad\addWithPreComma{#1}}
	\newcommand{\lReac}[1][]{Q_\sLoad\addWithPreComma{#1}}
	\newcommand{\lApp}{S_\sLoad}
	\newcommand{\lActNom}{P_{0}}
	\newcommand{\lReacNom}{Q_{0}}

	\newcommand{\lYp}[1][]{Y_{\text{P}\addWithPreComma{#1}}}
	\newcommand{\lIp}[1][]{I_{\text{P}\addWithPreComma{#1}}}
	\newcommand{\lPp}[1][]{P_{\text{P}\addWithPreComma{#1}}}
	\newcommand{\lZpCoeff}[1][]{\denotePolyLoadCoeff_{Z,\text{P}\addWithPreComma{#1}}}
	\newcommand{\lIpCoeff}[1][]{\denotePolyLoadCoeff_{I,\text{P}\addWithPreComma{#1}}}
	\newcommand{\lPpCoeff}[1][]{\denotePolyLoadCoeff_{P,\text{P}\addWithPreComma{#1}}}

	\newcommand{\lYq}[1][]{Y_{\text{Q}\addWithPreComma{#1}}}
	\newcommand{\lIq}[1][]{I_{\text{Q}\addWithPreComma{#1}}}
	\newcommand{\lPq}[1][]{P_{\text{Q}\addWithPreComma{#1}}}
	\newcommand{\lZqCoeff}[1][]{\denotePolyLoadCoeff_{Z,\text{Q}\addWithPreComma{#1}}}
	\newcommand{\lIqCoeff}[1][]{\denotePolyLoadCoeff_{I,\text{Q}\addWithPreComma{#1}}}
	\newcommand{\lPqCoeff}[1][]{\denotePolyLoadCoeff_{P,\text{Q}\addWithPreComma{#1}}}
	
	\newcommand{\lNp}[1][]{n_{\text{P}\addWithPreComma{#1}}}
	\newcommand{\lNq}[1][]{n_{\text{Q}\addWithPreComma{#1}}}
	
	\newcommand{\Cti}{\ensuremath{C_{\text{t}i}}}
	\newcommand{\Lti}{\ensuremath{L_{\text{t}i}}}

	\newcommand{\Lij}{\ensuremath{L_{ij}}}

\graphicspath{{./03_Img/}}

\begin{document}
	\selectlanguage{USenglish}

	\begin{frontmatter}

\title{Passivity Conditions for Plug-and-Play Operation of Nonlinear Static AC Loads}

\author[First]{Felix Strehle}
\author[First]{Albertus J. Malan}
\author[First]{Stefan Krebs}
\author[First]{S\"oren Hohmann}

\address[First]{Karlsruhe Institute of Technology (KIT), \\
	Kaiserstra{\ss}e 12, 76131 Karlsruhe, Germany (e-mail: \{albertus.malan,felix.strehle,stefan.krebs,soeren.hohmann\}@kit.edu).}

\begin{abstract} % Abstract of not more than 250 words.
: The complexity arising in AC microgrids from multiple interacting \emph{distributed generation units} (DGUs) with intermittent supply behavior requires local \emph{voltage-source inverters} (VSIs) to be controlled in a distributed or decentralized manner at primary level.
In \citep{Strehle19}, we use passivity theory to design decentralized, plug-and-play voltage and frequency controllers for such VSIs. However, the stability analysis of the closed-loop system requires a load-connected topology, in contrast to real grids where loads are arbitrarily located. In this paper, we expand our former approach by considering the more realistic and general case of nonlinear static AC loads (ZIP and exponential) at arbitrary locations within an AC microgrid. Investigating the monotonicity of differentiable mappings, we derive sufficient inequality conditions for the strict passivity of these nonlinear static AC loads.
Together with our plug-and-play VSI controller, this allows us to use passivity arguments to infer asymptotic voltage and frequency stability for AC microgrids with arbitrary topologies. An illustrative simulation validating our theoretical findings concludes our work.

\end{abstract}


\end{frontmatter}
	\section{Introduction} \label{sec:Introduction}

% General statement complexity
Inverter-based AC microgrids comprise electrical distribution networks linking a multitude of loads and various renewable energy sources, referred to as \emph{distributed generation units} (DGUs), which also include storage devices \citep{Lasseter01}\citep{Schiffer16}\citep{Olivares14}. 
When compared to classical power systems, voltage and frequency control in such AC microgrids represents a far more complex challenge which is the subject of much recent research activity.

% Why decentralized
Particularly the heavily increased number of interacting components (loads, DGUs) together with their intermittent supply/demand behavior requires scalability of the frequency and voltage control methods for the local VSIs at primary level \citep{Lasseter01}\citep{Guerrero13}.
Decentralized control methods enable such scalability as they only rely on local DGU information and measurements for the corresponding local VSI control design. This allows for the addition or removal of DGUs in a \emph{plug-and-play} fashion without requiring changes to any existing local controllers.
% Passivity suited for decentralized
An extensive framework for modular modeling and decentralized control methods of complex large-scale systems is passivity theory \citep{vdS17} \citep{Duindam09}\citep{Fiaz13}\citep{vdS16}\citep{Strehle18}. Thus, recent publications focusing on decentralized plug-and-play voltage and frequency controllers of VSIs in AC microgrids, use passivity-based approaches \citep{Nahata19}\citep{Strehle19}. Exploiting the compositional property of passive systems, such approaches provide a modular analysis of microgrid-wide voltage and frequency stability via the passivity of all subsystems within an AC microgrid and their passivity-preserving interconnection. 

%inadequately represents realistic situation in which pure load nodes can occur at arbitrary positions

% Current status drawback of passivity-basec PnP controllers for AC MGs
However, these former results are either (i) obtained under a load-connected AC microgrid topology in which all loads are connected to the controllable VSI terminals of DGUs, possibly by Kron reduction, \citep{Strehle19}, or (ii) necessitate solving linear matrix inequalities for each DGU plug-in/out operation \citep{Nahata19}. While the load-connected topology in (i) distorts the original topology and thus hampers insight into operationally relevant phenomena, e.g. line congestion or voltage transients at nodes that are reduced, the numerical optimization in (ii) might be infeasible and prevent the plug-in/out of the respective DGU.
Furthermore, both approaches use reduced forms of the nonlinear static ZIP load model (ZP in \cite{Strehle19}; ZI in \cite{Nahata19}). However, prevalent nonlinear static AC load models comprise full ZIP and exponential load models \cite[pp.~111-112]{Machowski08}\cite[pp.~33-34]{Canizares18}.

%\fs{Should this somehow be added? I think it disrupts focus and storyline: In DC microgrids with load-connected topology, full ZIP models have been incorporate into plug-and-play approaches. However, besides a load-connected topology, the DC case is considerably simplified as the static voltage-dependent load current function is scalar. In the AC case with balanced, three-phase signals, the resulting representation in \dq coordinates results in a two-dimensional vector function with coupled coordinates.}

% Contribution
In this paper, we address these issues by considering the more general case of nonlinear static ZIP and exponential AC loads at arbitrary locations within an AC microgrid.  Note that load-connected topologies are still included in that consideration as a special case.
%Layout
In a first step, we set up a port-Hamiltonian model of uncontrollable, lone-standing load nodes and include it in the modular, passivity-based voltage and frequency stability analysis from \cite{Strehle19}. In so doing, we expand our former results to arbitrary AC microgrid topologies comprising electrical lines, controllable DGU nodes, with optional uncontrollable local loads, and uncontrollable lone-standing load nodes. 
In order to investigate the passivity of the load nodes, we then establish a helpful reformulation of the load models as voltage-dependent current functions in \dq coordinates. To the best of the authors knowledge, such a representation is not directly available in the pertinent literature, where loads are commonly described by their consumed active and reactive powers (see for example \cite[pp.~111-112]{Machowski08}\cite[pp.~33-34]{Canizares18}). With this, we finally derive sufficient inequality conditions for the strict passivity of nonlinear static ZIP and exponential AC load models by investigating the monotonicity of the load current-voltage relation.
Additionally, we give an interesting perspective on how our results are related to incremental passivity of memoryless functions.

% Summary main contributions
In summary, our main contributions comprise (i) the extension of our decentralized plug-and-play approach from \cite{Strehle19} by uncontrollable lone-standing load nodes to guarantee asymptotic voltage and frequency stability of AC microgrids with arbitrary topologies; (ii) the proposition of sufficient inequality conditions for the strict passivity of the prevalent nonlinear static ZIP and exponential AC load models.
	\section{Problem Formulation} \label{sec:Problem}
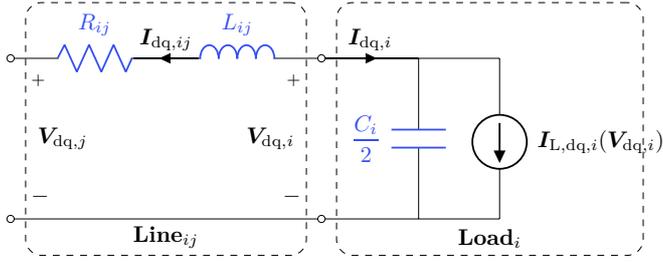
\begin{figure}[!t]
	\centering
	\resizebox{\columnwidth}{!}{%
		\tikzsetnextfilename{03_Img/load_circuit_model}%
		\begin{tikzpicture}
	\def\cHeight{2.5cm}		% Circuit height
	
	\coordinate(port_left_low);
%%%%%%%% Line Circuit	
	\draw
		% Draw Ports and RL branches
		(port_left_low) to [open] ++(0,\cHeight) coordinate(port_left_high)
		to [short, o-] ++(0.7, 0) coordinate(line_top_left)
		to [R, color=colorLineComp, colorLineComp, l=${\cR[ij]}$] ++(1.2,0)
		to [short, i^<=${\Idq[ij]}$] ++(1.0, 0)
		to [L, color=colorLineComp, colorLineComp, l=${\cL[ij]}$] ++(1.2,0) coordinate(line_top_right)
		to [short] ++(0.7,0) coordinate(port_right_high)
		to [open] ++(0,-\cHeight) coordinate(port_right_low)
		to [short, -o] (port_left_low)
		
		(line_top_right) to [open] ++(0,-\cHeight) coordinate(line_bottom_right)

		% Draw C branch
		(port_right_high) to [short,o-,i^>=${\IPdq[i]}$] ++(1.5,0) coordinate(line_C_high)
		to [C, color=colorLineComp, colorLineComp, l_=$\dfrac{\cC[i]}{2}$] ++(0,-\cHeight) coordinate(line_C_low)
		to [short,-o] (port_right_low)
		
		% Draw voltages
		(line_top_left) to [open] ++(-0.5,0)
		to [open, v^>=${\VLdq[j]}$] ++(0, -\cHeight)
		(line_top_right) to [open] ++(0.5,0)
		to [open, v_>=${\VLdq[i]}$] ++(0, -\cHeight);
	
%%%%%%%% Load Circuit
	\draw
		% Load (Current Source) Branch
		(line_C_high) to [short] ++(1.25cm,0) coordinate(load_top)
		to [short] ++(0,-0.1cm)
		to [american current source, i>^=${\ILdq[i](\VLdq[i])}$] ++(0,{-\cHeight+0.1cm}) coordinate(load_bottom)
		to [short] (line_C_low);
%%%%%%%% End DGU Circuit

%%%%%%%% Draw bounding boxes
	% System bounding box
	\path (line_top_left) +(-0.35,0.75) coordinate (line_box_top_left);
	\path (line_bottom_right) +(0.35,-0.45) coordinate (line_box_bottom_right);
	\path (load_bottom) +(2.1,-0.45) coordinate (load_box_bottom_right);
	\path (port_right_high) +(0.35,0.75) coordinate (load_box_top_left);
	
	\begin{scope}[on background layer]
		% Draw Load box
		\node[draw,dashed,rounded corners=0.25cm,fit=(line_box_bottom_right) (line_box_top_left)] (line_box) {};
		\node[above] at (line_box.south) {\textbf{Line$_{ij}$}};
		\node[draw,dashed,rounded corners=0.25cm,fit=(load_box_bottom_right) (load_box_top_left)] (load_box) {};
		\node[above] at (load_box.south) {\textbf{Load$_{i}$}};
	\end{scope}
%%%%%%%% End Draw boundig boxes
		
\end{tikzpicture}%
	}
	%	\scalebox{1.0}{\inputtikz{03_Img/load_circuit_model}}
	\caption{Circuit diagram of a three-phase $\pi$-line (in \textbf{\color{colorLineComp}blue}) connected to a nonlinear static AC load}
	\label{fig:load_circuit_model}
\end{figure}

\begin{figure} 
	\centering
	\scalebox{.75}{\begin{tikzpicture}
	\newcommand{\dguTextDGU}[1]{DGU$_{#1}$}
	\newcommand{\dguTextLoad}[1]{Load$_{#1}$}
	\newcommand{\lineText}[3]{Line$_{#1,#2}$}

	\def\dguXdist{4.5cm}
	\def\dguYdist{3.5cm}
	\def\dguNodes{2,3,5}
	\def\loadNodes{1,4}
	\def\lineNodes{1/2, 1/3, 2/3, 2/4, 3/4, 3/5, 4/5}
	\def\lineNodesLengths{1/2/3, 1/3/7, 2/3/5, 2/4/8, 3/4/5, 3/5/2, 4/5/4}
	
	\def\lineNodesFilled{1/2, 1/3, 2/3, 2/4, 3/4}
	\def\lineNodesDashed{3/5, 4/5}
	
	\tikzstyle{dgu}    = [draw, rectangle split, rounded corners, rectangle split parts=2, rectangle split part fill={red!30, orange!30}]
	\tikzstyle{load}   = [draw, rectangle,rounded corners=1.5mm, minimum height = 1.1cm, fill={orange!30}]
	\tikzstyle{line}   = [draw, rounded rectangle, minimum height = 0.6cm, fill={blue!20},text width=1cm,align=center]
	\tikzstyle{arrowInFilled}  = [-latex, line width=1.0pt]
	\tikzstyle{arrowOutFilled} = [-latex, line width=1.0pt]
	\tikzstyle{arrowInDashed}    = [-latex, dotted, line width=1.2pt]
	\tikzstyle{arrowOutDashed}   = [-latex, dotted, line width=1.2pt]
%	\tikzstyle{arrowInFilled}  = [line width=1.2pt]
%	\tikzstyle{arrowOutFilled} = [line width=1.2pt]
%	\tikzstyle{arrowInDashed}    = [dotted, line width=1.2pt]
%	\tikzstyle{arrowOutDashed}   = [dotted, line width=1.2pt]
	
	% Define DGU coordinates
	\coordinate(cDgu1);
	\path (cDgu1) +(0,\dguYdist) coordinate (cDgu2);
	\path (cDgu1) +(\dguXdist,0) coordinate (cDgu3);
	\path (cDgu2) +(\dguXdist,0) coordinate (cDgu4);
	\path (cDgu3) +(\dguXdist,\dguYdist / 2) coordinate (cDgu5);
	
	% Define Line coordinates
	\foreach \a/\b in \lineNodes {\coordinate (cLine\a\b) at ($(cDgu\a)!0.5!(cDgu\b)$);}
	
	% Draw DGU nodes
	\foreach \a in \dguNodes {\node[dgu](dgu\a) at(cDgu\a) {\dguTextDGU{\a}\nodepart{second}\dguTextLoad{\a}}; }
	
	% Draw Load nodes
	\foreach \a in \loadNodes {\node[load](dgu\a) at(cDgu\a) {\dguTextLoad{\a}}; }

	% Draw Line nodes
	\foreach \a/\b/\l in \lineNodesLengths {\node[line](line\a\b) at(cLine\a\b){\lineText{\a}{\b}{\l}};}
	
	% Draw Filled Arrows
	\foreach \a/\b in \lineNodesFilled {
		\draw[arrowInFilled](dgu\a) to (line\a\b);
		\draw[arrowOutFilled] (line\a\b) to (dgu\b);	}
	
	% Draw Dashed Arrows
	\foreach \a/\b in \lineNodesDashed {
		\draw[arrowInDashed](dgu\a) to (line\a\b);
		\draw[arrowOutDashed] (line\a\b) to (dgu\b);	}
	
%	%%%%% Alternative Arrows and Lines
%	% Draw Filled Arrows
%	\foreach \a/\b in \lineNodesFilled {
%		\draw[arrowInFilled](dgu\a) to (dgu\b);	}
%	
%	% Draw Dashed Arrows
%	\foreach \a/\b in \lineNodesDashed {
%		\draw[arrowInDashed](dgu\a) to (dgu\b);	}
%	
%	% Draw Line nodes
%	\foreach \a/\b/\l in \lineNodesLengths {\node[line](line\a\b) at(cLine\a\b){\lineText{\a}{\b}{\l}};}
\end{tikzpicture}}
	\caption{%
		Bipartite graph representation of an AC microgrid with electrical lines interconnecting controllable DGU nodes (with optional local loads) and uncontrollable load nodes; dotted lines indicate the plug-and-play nature of the microgrid}
	\label{fig:microgrid_bipartite}
\end{figure}
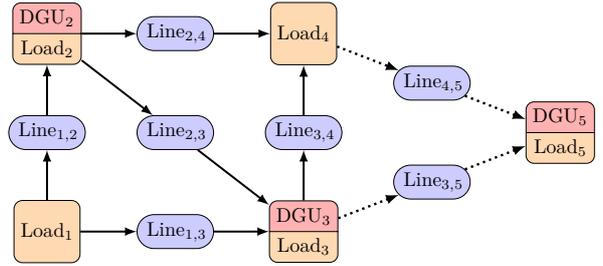

The overarching problem we investigate in this work is the asymptotic voltage and frequency stability of AC microgrids with arbitrary topologies. The AC microgrid is modeled as in \cite{Strehle19} with electrical lines, controllable DGUs, with optional uncontrollable local loads, and uncontrollable lone-standing loads in the \emph{dq} reference frame rotating at $\freqRef = 2 \pi \, \SI{50}{\hertz}$. The zero-sequence is neglected by assuming a balanced network. As we are considering nonlinear static AC load models, we model the lone-standing loads as voltage-dependent current sources connected via $\pi$-model electrical lines to the remaining AC microgrid (see \autoref{fig:load_circuit_model}). Consequently, an AC microgrid can be represented by a bipartite graph as in \autoref{fig:microgrid_bipartite}.
Without loss of generality, we consider only the connection of the load with a single line in the subsequent sections. 
The generalization to multiple connecting lines can readily be obtained by replacing $\cC[i]$ with the sum of all parallel-connected capacitances $\cC[ij]$ of these lines and $\IPdq[i]$ with the negative sum of all outgoing $\ILinedq[ij]$.

\subsection{Preliminaries}
For the remaining part of this contribution, we establish the following assumption which holds under normal grid conditions:
\begin{assumption} \label{assumption:v>0andloadparameters>=0}
	Any voltages (node, reference, nominal) not in the \dq frame
	are strictly positive, i.e. $\Vamp(t) > 0$ for all $t \geq 0$. The reference frequency is strictly positive $\freqRef>0$. All parameters used in the load models are positive, i.e. real numbers greater or equal to zero.	
\end{assumption}
\begin{remark}
	Note that due to \autoref{assumption:v>0andloadparameters>=0}, we only use the term asymptotic voltage and frequency stability in this work and drop the denotation global. Although, from a practical perspective, the obtained stability result is global in the sense that it holds for the complete operationally relevant area of $\Vamp(t) > 0$ and $\freqRef>0$ for all $t \geq 0$.
\end{remark}
Furthermore we notice that variables in a \dq frame (rotating at frequency $\freqRef$) can be considered Cartesian representations of complex vectors with the equivalent polar representation $ \Vamp(t) \angle \theta(t) $. Asymptotically stable \dq systems imply $\displaystyle \lim\limits_{t \to \infty} \theta(t)=const.$, and are therefore also asymptotically stable as phasors with the frequency of the rotating phasors $ \omega \;{=}\;  \dot{\theta} + \freqRef \;{=}\; \freqRef$, since $\dot{\theta} \;{=}\;  0$.
Asymptotic frequency stability is therefore implicitly present in an asymptotically stable \dq system. This means that microgrid-wide asymptotic voltage stability in \dq coordinates implies microgrid-wide asymptotic frequency stability, if the various local \dq frames at each node are synchronized. Thus, we establish the following Assumption similar to \citep[Assumption~2]{Cucuzzella18} \cite[Remark~1]{Nahata19}:
\begin{assumption} \label{assumption:dq_synchrony}
	The various \dq reference frames in the local VSI controllers and at load nodes are synchronized. 
\end{assumption}
However, note that the \dq reference frames at load nodes are only used as means to facilitate the analysis of the balanced, three-phase AC signals. There are no actual controller clocks to be synchronized at such nodes.

\subsection{Voltage and Frequency Stability}
Investigating the asymptotic voltage and frequency stability of AC microgrids with arbitrary topologies means that in addition to the asymptotic stability of controllable DGU equilibria (cf.\ (10) in \cite{Strehle19})
\begin{equation}\label{eq:Stability:dgu_equilibrium}
\xRef[i] = \left[\Lti \IdRef[i], \, \Lti \IqRef[i], \, \Cti \VdRef[i], \, \Cti \VqRef[i]\right]^\Transpose,
\end{equation}
we now also investigate the asymptotic stability of uncontrollable load equilibria. For this, we model the load node $i$ in \autoref{fig:load_circuit_model} as a port-Hamiltonian system with a nonlinear resistive structure (cf. \cite[p.~114]{vdS17})
\begin{subequations} \label{eq:Problem:loadphs}
	\begin{align}
		\begin{bmatrix}
			\cC[i] \Vddot[i]\\
			\cC[i] \Vqdot[i]
		\end{bmatrix}&=
		\JL[i]
		\begin{bmatrix}
			\Vd[i]\\
			\Vq[i]
		\end{bmatrix}-\ILdq[i](\Vdq[i])+
		\KL[i]
		\begin{bmatrix}
			\IPd[i]\\
			\IPq[i]
		\end{bmatrix} \nonumber\\
		\z[i]&=\KL[i]^\Transpose
		\begin{bmatrix}
			\Vd[i]\\
			\Vq[i]
		\end{bmatrix}\\
		\HLx[i]&=\frac{1}{2} \xL[i]^\Transpose \text{Diag}\left[\frac{1}{\cC[i]}, \frac{1}{\cC[i]} \right]  \xL[i] \nonumber 
	\end{align}
	with states $\xL[i]=\left[\cC[i] \Vd[i], \cC[i] \Vd[i] \right]^\Transpose$, uncontrollable input $\vd[i]=\left[\IPd[i], \IPq[i] \right]^\Transpose$, uncontrollable output $\z[i]=\left[\Vd[i], \Vq[i] \right]^\Transpose$, nonlinear resistive structure
		\begin{equation}\label{eq:Problem:nonlinear_R}
		\mathcal{R}(\xL[i])=\ILdq[i](\Vdq[i]),
		\end{equation}
	and interconnection and uncontrollable input matrices
	\begin{equation}\label{eq:Modeling:phs_matrices}
	\JL[i]=\begin{bmatrix}
	0				&	\freqRef\cC[i]\\
	-\freqRef\cC[i]	&	0
	\end{bmatrix}, \quad
	\KL[i]=\begin{bmatrix}
	1	&	0\\
	0	&	1
	\end{bmatrix}.
	\end{equation}
\end{subequations}
As load node voltages $\Vdq[i]$ are not directly controllable, the equilibrium
\begin{equation} \label{eq:Problem:load_equilibrium}
\xLRef[i] = \cC[i] \VdqRef[i],
\end{equation}
of \eqref{eq:Problem:loadphs} is specified by exchange current $\vd[i]=\Idq[i]$ resulting from the voltages $\Vdq[j]$ at neighboring nodes (see \autoref{fig:load_circuit_model}).
For the stability analysis of \eqref{eq:Problem:load_equilibrium}, we define the error variables
\begin{align} \label{eq:Problem:error_variable_voltage}
	\vecError{\Vdq[i]} &\coloneqq \Vdq[i] - \VdqRef[i],& \vecError{\Vdq[i]}&\in \Reals^2, \\
	\stateErrorL[i] &\coloneqq \xL[i] - \xLRef[i],& \stateErrorL[i] &\in \Reals^2,
\end{align}
and set up the error system of the load model \eqref{eq:Problem:loadphs} as
\begin{subequations} \label{eq:errorsystem}
	\begin{align}
		\cC[i]\dVecError{\Vdq[i]}&=
		\JL[i]
		\vecError{\Vdq[i]}
		- \left( \ILdq[i](\Vdq[i])-\ILdq[i](\VdqRef[i])\right) \nonumber\\
		&\dots +\KL[i] \left( \vd[i]-\vdRef[i]\right),\\
		\z[i]-\zRef[i]&=\KL[i]^\Transpose \vecError{\Vdq[i]},\\
		\label{eq:errorsystem:Hamiltonian}
		\HLe[i] &= \frac{1}{2} \, \stateErrorL[i]^\Transpose \; \text{Diag}\left[\frac{1}{\cC[i]}, \frac{1}{\cC[i]} \right] \stateErrorL[i]
	\end{align}
\end{subequations}
%
%with \eqref{eq:Modeling:current_ZIP} and \eqref{eq:Modeling:current_Exp} for ZIP and exponential loads, respectively. 
%
\begin{remark}
	Note that the current references $\Lti \IdRef[i], \Lti \IqRef[i]$ in \eqref{eq:Stability:dgu_equilibrium} as well as the line equilibria (cf.\ (13) in \cite{Strehle19})
	\begin{equation} \label{eq:Stability:line_equilibrium}
	\xRef[ij] = \left[\Lij\IdRef[ij], \, \Lij \IqRef[ij]\right]^\Transpose
	\end{equation}
	are not specified explicitly and follow as a consequence of the load demand and  node voltages.
\end{remark}
With \eqref{eq:errorsystem}, we can formulate the following extended version of our former Proposition~7 from \cite{Strehle19}:

\begin{proposition} \label{prop:microgrid_passivity}
	An AC microgrid with an arbitrary topology represented by a bipartite graph as in \autoref{fig:microgrid_bipartite} consisting of $\pi$-model lines, strictly passive DGU PHSs, and strictly passive nonlinear static ZIP and exponential loads described by \eqref{eq:errorsystem} is itself strictly passive. Its asymptotically stable equilibrium $\xMRef[]$ is given by the combined equilibria $\xRef[ij]$, $\xRef[i]$, and $\xLRef[i]$ of the individual subsystems.
\end{proposition}
\begin{pf}
	From \cite{Strehle19} we know that $\pi$-model lines and DGUs with our plug-and-play controller both are strictly passive systems. The minima of their storage functions (Hamiltonians) are $\xRef[ij]$ and $\xRef[i]$, respectively. The equilibria of the load nodes are the minima of \eqref{eq:errorsystem:Hamiltonian} which is $\stateErrorL[i]=0$ implying \eqref{eq:Problem:load_equilibrium} and thus the unknown-steady voltages $\VdqRef[i]$. If the load model error system \eqref{eq:errorsystem} is strictly passive for ZIP and exponential loads, respectively, then an AC microgrid with arbitrary topology comprises only strictly passive subsystems. 
%	Their interconnection is established via ideal flow (current) constraints which are power-preserving Dirac structures \cite[p.~100]{Duindam09}. Their equilibria specified by the minima of \eqref{eq:errorsystem:Hamiltonian} are given by $\stateErrorL[i]=0$ implying \eqref{eq:Stability:load_equilibrium} and thus the unknown-steady voltages $\VdqRef[i]$ at pure load nodes. 
	The remainder of the proof follows according to \cite{Strehle19} via the interconnection of passive systems and Lyapunov's direct method.  \qedsymbol
\end{pf}
In light of \autoref{prop:microgrid_passivity} and our former results, the voltage and frequency stability analysis of AC microgrids with arbitrary topologies reduces to investigating under which conditions the error system \eqref{eq:errorsystem} of nonlinear static ZIP and exponential AC loads is strictly passive. For this, we first establish the static load current-voltage relation $\ILdq[i](\Vdq[i])$.
	\section{Modeling of Nonlinear Static AC Loads}\label{sec:Modeling}
In order to obtain $\ILdq[i](\Vdq[i])$, we introduce standard AC load models described by the consumed active and reactive powers (\autoref{sec:Modeling:Loads}). Then, we derive the desired current equations in the \dq frame (\autoref{sec:Modeling:Load_I_V}).
\subsection{ZIP and Exponential AC Loads}\label{sec:Modeling:Loads}
The most common nonlinear static AC load models are polynomial and exponential models described by the active and reactive powers $\lAct(\Vamp)$ and $\lReac(\Vamp)$ as voltage-dependent functions \cite[pp.~111-112]{Machowski08}\cite[pp.~33-34]{Canizares18}\cite[pp.~95ff]{vanCutsem98}.
Polynomial models comprise constant impedance ($\denotePolyLoadCoeff_Z$), constant current ($\denotePolyLoadCoeff_I$) and constant power ($\denotePolyLoadCoeff_P$) coefficients, leading to the ZIP load equations
\begin{subequations} \label{eq:Modeling:zip_full}
	\begin{align}
		\lAct(\Vamp) &= \lActNom \left[\lZpCoeff \left(\frac{\Vamp}{\VampNominal}\right)^2 + \lIpCoeff \left(\frac{\Vamp}{\VampNominal}\right) + \lPpCoeff\right] \, , \\
		\lReac(\Vamp) &= \lReacNom \left[\lZqCoeff \left(\frac{\Vamp}{\VampNominal}\right)^2 + \lIqCoeff \left(\frac{\Vamp}{\VampNominal}\right) + \lPqCoeff\right] \, ,
	\end{align}
\end{subequations}
where $\VampNominal$ is the nominal phase-to-phase RMS value (e.g.\ \SI{400}{\volt}), and $\lActNom$ and $\lReacNom$ are the nominal active and reactive powers. By grouping the coefficients and nominal values in \eqref{eq:Modeling:zip_full} into the model parameters
\begin{equation}\label{eq:Modeling:zip_parameters}
	\begin{split}
		\lYp&=\frac{\lZpCoeff\lActNom}{\VampNominal^2},\
		\lIp=\frac{\lIpCoeff \lActNom}{\VampNominal^2},\
		\lPp=\frac{\lPpCoeff \lActNom}{\VampNominal^2},\\ \lYq&=\frac{\lZqCoeff\lReacNom}{\VampNominal^2},\
		\lIq=\frac{\lIqCoeff \lReacNom}{\VampNominal^2},\
		\lPq=\frac{\lPqCoeff \lReacNom}{\VampNominal^2},\
	\end{split}
\end{equation}
we obtain the simplified equations
\begin{subequations} \label{eq:Modeling:zip_simple}
	\begin{align}
		\lAct(\Vamp) &= \lYp \Vamp^2 + \lIp \Vamp + \lPp \, , \\
		\lReac(\Vamp) &= \lYq \Vamp^2 + \lIq \Vamp + \lPq \, .
	\end{align}
\end{subequations}
Note that the constant impedances (Z) are expressed as admittances (Y).
Exponential load models, on the other hand, are given by
\begin{subequations} \label{eq:Modeling:exp_full}
	\begin{align}
		\lAct(\Vamp) &= \lActNom \left(\frac{\Vamp}{\VampNominal}\right)^{\lNp} \, , \\
		\lReac(\Vamp) &= \lReacNom \left(\frac{\Vamp}{\VampNominal}\right)^{\lNq} \, ,
	\end{align}
\end{subequations}
where the model parameters are the voltage indexes $\lNp$ and $\lNq$ of the active and reactive power, respectively. 
%They offer a more generalized approach at the cost of the intuitive interpretability of ZIP loads.
%
\begin{remark} \label{remark:validityloads}
As per \cite[pp.~110-112]{Machowski08}, the models in \eqref{eq:Modeling:zip_full}, \eqref{eq:Modeling:zip_simple}, and \eqref{eq:Modeling:exp_full} are only accurate above $0.7 \, \VampNominal$. Below $0.7 \, \VampNominal$, real loads typically exhibit a rapid power drop and are only modeled as constant impedances (Z) or rather admittances (Y)
\begin{subequations} \label{eq:Modeling:z_model}
	\begin{align}
		\lAct(\Vamp) &= \lActNom \left[\lZpCoeff \left(\frac{\Vamp}{\VampNominal}\right)^2 \right]=\lYp \Vamp^2, \\
		\lReac(\Vamp) &= \lReacNom \left[\lZqCoeff \left(\frac{\Vamp}{\VampNominal}\right)^2 \right]=\lYq \Vamp^2.
	\end{align}
\end{subequations}
The combination of \eqref{eq:Modeling:z_model} for $\Vamp < 0.7 \, \VampNominal$ and \eqref{eq:Modeling:zip_simple} and \eqref{eq:Modeling:exp_full}, respectively, for $\Vamp \geq 0.7 \, \VampNominal$ is referred to as two-tier load model. 
However, \eqref{eq:Modeling:z_model} is a special case of \eqref{eq:Modeling:zip_simple} for $\lIp, \lIq, \lPp, \lPq = 0$. Thus, for the sake of brevity, we focus on \eqref{eq:Modeling:zip_simple} and \eqref{eq:Modeling:exp_full} in the sequel.
\end{remark} 
\subsection{Power Conjugated Voltage-Current Functions of Loads} \label{sec:Modeling:Load_I_V}
Relating the power-conjugated input and output as $\ILdq[i](\Vdq[i])$ is achieved by considering the voltage-dependent powers $\lAct$ and $\lReac$ of the loads.
%%%%%%%%%%%%%%%%%%%%%%%%%%%%%%%%%%%%%%%%%%%%%%%%%%%%%%%%%%
\begin{lemma}[Current equations of a load] \label{lemma:Modeling:Load_current_eq}
	A balanced, nonlinear, static three-phase AC load described by voltage-dependent active and reactive power equations, $\lAct(\Vamp)$ and $\lReac(\Vamp)$, respectively, can be described by the \dq current equations
	\begin{equation} \label{eq:Load_Passivity:Load_currents:dq_currents}
	\ILdq(\VLdq) = \frac{1}{\Vamp^2} \begin{bmatrix}
	\lAct(\Vamp) & \lReac(\Vamp) \\
	-\lReac(\Vamp) & \lAct(\Vamp)
	\end{bmatrix} \begin{bmatrix}
	\VLd \\ \VLq
	\end{bmatrix} \, ,
	\end{equation}
	where $\Vamp$ is the amplitude (2-norm) of the \dq load voltage vector $\VLdq$, i.e.\
	\begin{equation}\label{eq:Load_Passivity:Load_currents:voltage_amp}
		\Vamp^2 = \VLd^2 + \VLq^2 \, .
	\end{equation}
\end{lemma}
\begin{pf}
	Let the instantaneous complex power of a load $\lApp$ be expressed in terms of \dq voltages $\VLdq$ and currents $\ILdq$ as in \cite{Schiffer16} \cite[p.~57]{vanCutsem98}
	\begin{align} 
		\lApp &= \lAct + \imag \lReac \nonumber \\
		\label{eq:Modeling:Load_currents:power_dq}
		&= \left(\VLd \ILd + \VLq \ILq \right) + \imag \left(\VLq \ILd - \VLd \ILq\right) \, .
	\end{align}
	By equating the real and imaginary parts of \eqref{eq:Modeling:Load_currents:power_dq}, we obtain the linear system
	\begin{equation}\label{eq:Modeling:Load_currents:power_dq_system}
	\begin{bmatrix}
	\lAct \\ \lReac
	\end{bmatrix} = \begin{bmatrix}
	\VLd & \VLq \\
	\VLq & -\VLd
	\end{bmatrix} \begin{bmatrix}
	\ILd \\ \ILq
	\end{bmatrix} \, .
	\end{equation}
	We then solve \eqref{eq:Modeling:Load_currents:power_dq_system} for $\ILdq$. This yields
	\begin{subequations} \label{eq:Modeling:Load_currents:dq_currents_calculated}
		\begin{align}
		\ILd(\VLdq) &= \frac{\VLd\lAct(\Vamp) + \VLq \lReac(\Vamp)}{\VLd^2 + \VLq^2} \, , \\
		\ILq(\VLdq) &= \frac{\VLq\lAct(\Vamp) - \VLd \lReac(\Vamp)}{\VLd^2 + \VLq^2} \, ,
		\end{align}
	\end{subequations}
	which is equivalent to \eqref{eq:Load_Passivity:Load_currents:dq_currents}. \qedsymbol
\end{pf}
%%%%%%%%%%%%%%%%%%%%%%%%%%%%%%%%%%%%%%%%%%%%%%%%%%%%%%%%%%%%%%%%%%%%
By applying \autoref{lemma:Modeling:Load_current_eq}, we calculate the power-conjugated voltage-current functions for the ZIP load in \eqref{eq:Modeling:zip_simple} as
\begin{equation} \label{eq:Modeling:current_ZIP}
	\underset{(2\times 1)}{\ILdq(\VLdq)} = \begin{bmatrix}
		\Bigg. \dfrac{\lPp \VLd + \lPq \VLq}{\Vamp^2} + \dfrac{\lIp \VLd + \lIq \VLq}{\Vamp} \\[-4pt]
		\Big. + \lYp \VLd + \lYq \VLq \\
		\hdashline
		\Bigg. \dfrac{\lPp \VLq - \lPq \VLd}{\Vamp^2} + \dfrac{\lIp \VLq - \lIq \VLd}{\Vamp} \\[-4pt]
		\Big. + \lYp \VLq - \lYq \VLd
	\end{bmatrix},
\end{equation}
and for the exponential load in \eqref{eq:Modeling:exp_full} as 
\begin{equation} \label{eq:Modeling:current_Exp}
	\underset{(2\times 1)}{\ILdq(\VLdq)} = \begin{bmatrix}
		\Bigg. \dfrac{\lActNom \Vamp^{\lNp - 2}}{\VampNominal^{\lNp}} \VLd + \lReacNom \dfrac{\Vamp^{\lNq - 2}}{\VampNominal^{\lNq}} \VLq \\
		\hdashline
		\Bigg. \dfrac{\lActNom \Vamp^{\lNp - 2}}{\VampNominal^{\lNp}} \VLq - \lReacNom \dfrac{\Vamp^{\lNq - 2}}{\VampNominal^{\lNq}} \VLd
	\end{bmatrix}.
\end{equation}
%
%and for the Z load below $0.7 \, \VampNominal$ in \eqref{eq:Modeling:z_model} as
%\begin{equation}\label{eq:Modeling:current_z}
%\underset{(2\times 1)}{\ILdq(\VLdq)} = \begin{bmatrix}
%\Big.\lYp \VLd + \lYq \VLq \\
%\hdashline
%\Big.\lYp \VLq - \lYq \VLd
%\end{bmatrix}.
%\end{equation}
	\section{Passivity of Nonlinear Static AC Loads} \label{sec:Load_Passivity}
In this section, we establish sufficient conditions for the strict passivity of the ZIP and exponential load model error systems \eqref{eq:errorsystem} with \eqref{eq:Modeling:current_ZIP} and \eqref{eq:Modeling:exp_full}, respectively. During the passivity analysis, we use the following established results from mathematics, which we repeat here for consistency:
\begin{lemma} \label{lemma:monotone_func}
	(Monotonicity of differentiable mappings)
	A differentiable mapping $\f{}{} \colon \Reals^m \rightarrow \Reals^m$ is monotone if and only if its Jacobian $\dfdu$ is positive semi-definite for all $\vu$. Furthermore, it is strictly monotone if $\dfdu$ is positive definite for all $\vu$. \cite[Prop.~ 12.3]{Rockafellar98}
\end{lemma}
\begin{lemma} \label{lemma:pos_def_symmetric_part}
(Positive definiteness of real quadratic matrices)
Any real quadratic matrix $\A\in\Reals^{n\times n}$, not necessarily symmetric, is positive (semi)-definite, if and only if its symmetric part $\A_{\text{s}}=\frac{1}{2}\left(\A+\A^\Transpose \right)$ is positive (semi)-definite. \cite[pp.~533--534]{Rockafellar98}
\end{lemma}
\begin{remark} \label{remark:drop_subscript_i}
	For clarity in the subsequent analysis, the subscript $i$ is dropped from all variables and parameters in this section, i.e.\ $\Vdq := \Vdq[i]$ etc.
\end{remark}
\subsection{Passivity conditions} \label{sec:Load_Passivity:Passivity}
\begin{proposition} \label{prop:passivity_conditions}
The load model error system \eqref{eq:errorsystem} is strictly passive, if
\begin{itemize}
	\item for ZIP load models with \eqref{eq:Modeling:current_ZIP} it holds that
	\begin{subequations} \label{eq:Load_Passivity:zip_restrictions}
		\begin{align} \label{eq:Load_Passivity:zip_restrictions:first}
			\lYp + \dfrac{\lIp}{2 \Vamp} &> 0,\\
			\lYp^2  \Vamp^4 + \lYp \lIp \Vamp^3 &> \frac{1}{4} \Iq \Vamp^2 + \left(\lIp \lPp + \lIq \lPq\right) \Vamp \dots \nonumber \\
			& \; \dots + \left(\lPp^2 +\lPq^2\right). \label{eq:Load_Passivity:zip_restrictions:second}
		\end{align}
	\end{subequations}
	\item for exponential load models with \eqref{eq:Modeling:current_Exp} it holds that
	\begin{subequations} \label{eq:Load_Passivity:exp_restrictions}
		\begin{align} \label{eq:Load_Passivity:exp_restrictions:first}
			\lNp \lActNom &> 0 , \\ 
			\label{eq:Load_Passivity:exp_restrictions:second}
			4\left(\lNp - 1\right) \lActNom^2 \left(\frac{\Vamp}{\VampNominal}\right)^{2 \lNp} &> \left(\lNq - 2\right)^2 \lReacNom^2 \left(\frac{\Vamp}{\VampNominal}\right)^{2 \lNq} 
		\end{align}
	\end{subequations}
\end{itemize}
\end{proposition}
\begin{pf}
The load model error system \eqref{eq:errorsystem} is strictly passive w.r.t. the supply rate $(\z[]-\zRef[])^\Transpose (\vd[]-\vdRef[])$ and the positive definite storage function \eqref{eq:errorsystem:Hamiltonian}, i.e. $\HLe[]: \Reals^2 \to \Reals^+, \HL{}{}{{\vec{0}}}=0, \HLe[]>0 \;\stateErrorL[]\neq \vec{0}$, if
\begin{multline}\label{eq:Load_Passivity:passivity_ineq}
	\HLedot[]=\vecError{\Vdq[]}^\Transpose \left(\JL \vecError{\Vdq[]} - \left( \ILdq[](\Vdq[])-\ILdq[](\VdqRef[])\right) \right)\\ + \vecError{\Vdq[]}^\Transpose \KL (\vd[]-\vdRef[])< (\z[]-\zRef[])^\Transpose (\vd[]-\vdRef[])
\end{multline}
holds. This is given if
\begin{equation}\label{eq:load_Passivity:passivity_ineq_simplified}
	\vecError{\Vdq[]}^\Transpose \left( \ILdq[](\Vdq[])-\ILdq[](\VdqRef[])\right)>0,
\end{equation}
which results with \eqref{eq:Problem:error_variable_voltage} in 
\begin{equation}\label{eq:Load_Analysis:monotonicity_ineq}
	\left( \Vdq[]-\VdqRef[] \right)^\Transpose \left( \ILdq[](\Vdq[])-\ILdq[](\VdqRef[])\right)>0.
\end{equation}
In order to derive conditions under which \eqref{eq:Load_Analysis:monotonicity_ineq} is fulfilled, we consider the load current function $\ILdq[](\Vdq[])$ as a differentiable mapping $\ILdq:\setU \subset \Reals^2 \to \setY \subset \Reals^2$. According to \cite[Def.~12.1]{Rockafellar98}, such a mapping $\ILdq[](\Vdq[])$ is called strictly monotone, if \eqref{eq:Load_Analysis:monotonicity_ineq} is fulfilled. Thus, investigating the strict monotonicity of $\ILdq[](\Vdq[])$ serves as proxy for investigating the strict passivity of \eqref{eq:errorsystem}. To determine whether $\ILdq[](\Vdq[])$ represents a strictly monotone mapping, we use \autoref{lemma:monotone_func} and \autoref{lemma:pos_def_symmetric_part}.

Consequently, we first derive the Jacobians
\begin{equation} \label{eq:Load_Passivity:load_Jacobian}
	\nabla \ILdq[](\Vdq[])= \dILdV = \begin{bmatrix}
		\Bigg.\dPartial{\ILd}{\VLd} & \dPartial{\ILd}{\VLq} \\
		\Bigg.\dPartial{\ILq}{\VLd} & \dPartial{\ILq}{\VLq} \\
	\end{bmatrix},
\end{equation}
and extract their symmetric part
\begin{equation}\label{eq:Load_Passivity:load_Jacobian_sym}
	\frac{\nabla \ILdq[](\Vdq[])+\nabla \ILdq[]^\Transpose(\Vdq[]) }{2} \eqqcolon \left[\begin{array}{@{\quad}c@{\qquad}c@{\quad}}
		\Big.a & b \\
		\Big.b & c \\
	\end{array}\right].
\end{equation}
For the ZIP current function in \eqref{eq:Modeling:current_ZIP}, \eqref{eq:Load_Passivity:load_Jacobian_sym} results in
\begin{subequations} \label{eq:Load_Passivity:zip_Jacobian_sym}
	\begin{align}
		a = & \, \lYp + \dfrac{\lIp \VLq^2 - \lIq \VLd \VLq}{\Vamp^3} \nonumber \\
		& \, - \dfrac{\lPp (\VLd^2 - \VLq^2) + 2 \lPq \VLd \VLq}{\Vamp^4} \\[6pt]
		b = & \, \dfrac{\lIq \VLd^2 - \lIp \VLd \VLq}{\Vamp^3} + \dfrac{\lPq (\VLd^2 - \VLq^2) - 2 \lPp \VLd \VLq}{\Vamp^4} \\[6pt]
			c = & \, \lYp + \dfrac{\lIp \VLd^2 + \lIq \VLd \VLq}{\Vamp^3} \nonumber \\
		& \, + \dfrac{\lPp (\VLd^2 - \VLq^2) + 2 \lPq \VLd \VLq}{\Vamp^4}
	\end{align}
\end{subequations}
Similarly, for the exponential current function in \eqref{eq:Modeling:current_Exp}, \eqref{eq:Load_Passivity:load_Jacobian_sym} is given by
\begin{subequations} \label{eq:Load_Passivity:exp_Jacobian_sym}
	\begin{align}
		a = & \,\dfrac{\lActNom \left[(\lNp - 1) \VLd^2 + \VLq^2\right]}{\VampNominal^{\lNp}\Vamp^{4-\lNp}} +\dfrac{(\lNq - 2) \lReacNom \VLd \VLq}{\VampNominal^{\lNq}\Vamp^{4-\lNq}} \\[6pt]
		b = & \, \dfrac{(\lNp - 2) \lActNom \VLd \VLq}{\VampNominal^{\lNp}\Vamp^{4-\lNp}} + \dfrac{(\lNq - 2)\lReacNom \left[\VLd^2 - \VLq^2\right]}{2\VampNominal^{\lNq}\Vamp^{4-\lNq}} \\[6pt]
		c = & \, \dfrac{\lActNom \left[\VLd^2 + (\lNp - 1)\VLq^2\right]}{\VampNominal^{\lNp}\Vamp^{4-\lNp}} - \dfrac{(\lNq - 2) \lReacNom \VLd \VLq}{\VampNominal^{\lNq}\Vamp^{4-\lNq}}
	\end{align}
\end{subequations}
Note that \eqref{eq:Load_Passivity:Load_currents:voltage_amp} must be taken into account when determining the respective Jacobians. 

Then, we find the eigenvalues $\eig[1,2]$ of \eqref{eq:Load_Passivity:load_Jacobian_sym} with \eqref{eq:Load_Passivity:zip_Jacobian_sym} and \eqref{eq:Load_Passivity:exp_Jacobian_sym} using computer algebra software. For ZIP loads with \eqref{eq:Load_Passivity:load_Jacobian_sym} and \eqref{eq:Load_Passivity:zip_Jacobian_sym}, this results in
%
%\begin{align} \label{eq:Load_Passivity:zip_eigenvalues}
%	\eig[1,2] =& \lYp + \frac{\lIp}{2 \Vamp} \pm \frac{1}{\Vamp^2} \left[- \lZq^2 \Vamp^4 - \lIq \lZq \Vamp^3 + \frac{1}{4} \lIp^2 \Vamp^2 \Bigg.\right. \nonumber \\
%	&\left.+ \left(\lIp \lPp + \lIq \lPq\right) \Vamp + \left(\lPp^2 + \lPq^2\right) \Bigg.\right]^{\dfrac{1}{2}}
%\end{align}
\begin{multline} \label{eq:Load_Passivity:zip_eigenvalues_sym}
	\eig[1,2] = \lYp + \frac{\lIp}{2 \Vamp} \pm \frac{1}{\Vamp^2} \left[ \frac{1}{4} (\lIp^2+\lIq^2) \Vamp^2  \Bigg.\right. \\
	\left. \dots+ \left(\lIp \lPp + \lIq \lPq\right) \Vamp + \left(\lPp^2 + \lPq^2\right) \Bigg.\right]^{\frac{1}{2}} \, ,
\end{multline}
and for exponential loads \eqref{eq:Load_Passivity:load_Jacobian_sym} and \eqref{eq:Load_Passivity:exp_Jacobian_sym} in
%
%\begin{equation} \label{eq:Load_Passivity:exp_eigenvalues}
%	\eig[1,2] = \frac{1}{2 \Vamp^2}\left([\lNp \lActNom \left(\frac{\Vamp}{\VampNominal}\right)^{\lNp} \pm \sqrt{\left(\lNp - 2\right)^2 \lActNom^2 \left(\frac{\Vamp}{\VampNominal}\right)^{2 \lNp} - 4 \left(\lNq - 1\right) \lReacNom^2 \left(\frac{\Vamp}{\VampNominal}\right)^{2 \lNq}}\right)
%\end{equation}
\begin{multline} \label{eq:Load_Passivity:exp_eigenvalues_sym}
	\eig[1,2] = \frac{1}{2 \Vamp^2}\left(\lNp \lActNom \left(\frac{\Vamp}{\VampNominal}\right)^{\lNp} \pm \left[\left(\lNp - 2\right)^2 \lActNom^2  \left(\frac{\Vamp}{\VampNominal}\right)^{2 \lNp} \Bigg.\right.\right.\\
	\left.\left.\Bigg. \dots + \left(\lNq - 2\right)^2 \lReacNom^2 \left(\frac{\Vamp}{\VampNominal}\right)^{2 \lNq}\right]^{\frac{1}{2}}\right).
\end{multline}

Finally, we evaluate the conditions under which \eqref{eq:Load_Passivity:zip_eigenvalues_sym} and \eqref{eq:Load_Passivity:exp_eigenvalues_sym} are positive to infer positive definiteness of \eqref{eq:Load_Passivity:load_Jacobian_sym} with \eqref{eq:Load_Passivity:zip_Jacobian_sym} and \eqref{eq:Load_Passivity:exp_Jacobian_sym}, respectively. As we are considering eigenvalues of symmetric matrices, their values are always real \cite[p.~8]{Rugh96}. Consequently, the $\eig[2]$, where the square root terms are subtracted, are critical when investigating the positiveness of \eqref{eq:Load_Passivity:zip_eigenvalues_sym} and \eqref{eq:Load_Passivity:exp_eigenvalues_sym}. With this preliminary consideration and \autoref{assumption:v>0andloadparameters>=0}, we obtain \eqref{eq:Load_Passivity:zip_restrictions:first} and
\begin{multline}\label{eq:Load_Passivity:zip_restrictions_derivation}
	\lYp \Vamp^2 + \frac{\lIp}{2} \Vamp > \left[ \frac{1}{4} (\lIp^2+\lIq^2) \Vamp^2 + \left(\lIp \lPp + \lIq \lPq\right) \Vamp  \dots \Bigg.\right. \\
	\left. \dots + \left(\lPp^2 + \lPq^2\right) \Bigg.\right]^{\frac{1}{2}}
\end{multline}
from \eqref{eq:Load_Passivity:zip_eigenvalues_sym}, and \eqref{eq:Load_Passivity:exp_restrictions:first} and
\begin{multline}\label{eq:Load_Passivity:exp_restrictions_derivation}
	\lNp \lActNom \left(\frac{\Vamp}{\VampNominal}\right)^{\lNp}>\left[\left(\lNp - 2\right)^2 \lActNom^2  \left(\frac{\Vamp}{\VampNominal}\right)^{2 \lNp} \dots \Bigg.\right.\\
	\left.\Bigg. \dots + \left(\lNq - 2\right)^2 \lReacNom^2 \left(\frac{\Vamp}{\VampNominal}\right)^{2 \lNq}\Bigg.\right]^{\frac{1}{2}}
\end{multline}
from \eqref{eq:Load_Passivity:exp_eigenvalues_sym}. Since both sides in \eqref{eq:Load_Passivity:zip_restrictions_derivation} and \eqref{eq:Load_Passivity:exp_restrictions_derivation} are positive, by squaring them, we arrive at \eqref{eq:Load_Passivity:zip_restrictions:second} and \eqref{eq:Load_Passivity:exp_restrictions:second}, respectively. \qedsymbol
\end{pf}
\begin{remark}
	For $\Vamp < 0.7 \, \VampNominal$, we insert $\lIp, \lIq, \lPp, \lPq = 0$ (cf.\ \autoref{remark:validityloads}) in \eqref{eq:Load_Passivity:zip_restrictions} to obtain 
	\begin{equation}
		\lYp >0
	\end{equation} 
	as sufficient condition for the strict passivity of \eqref{eq:errorsystem}. 
\end{remark}
\begin{remark}
	Equation \eqref{eq:Load_Analysis:monotonicity_ineq} requires that the product of the power-conjugated input (voltage difference) and output (current difference) be positive. This corresponds to the incremental passivity \cite[p.~95]{vdS17} of static systems, i.e.\ memoryless systems with zero storage functions (cf. \cite[p.~228]{Khalil02} for passivity of memoryless functions). This in turn comes full circle to the monotonicity of the resistive relation specified by the voltage-dependent load current function $\ILdq[i](\VLdq[i])$ \cite[p.~92]{vdS14}. When considering passivity as a special case of dissipativity with specific quadratic supply rate, a similar link can be made to equilibrium-independent dissipativity of nonlinearities (cf.\ (10) in \cite{SimpsonPorco19}).
\end{remark}
	\section{Simulation} \label{sec:Simulation}
%
% 1) Goal of Simulation: What is to be demonstrated?
%	- Goal: demonstrate plausibility of results, passivity boundary of loads
%	- Show that load results in instability when passivity condition is not met
% 2) Description of the simulation setup: (Testsystem, events, inputsignals, parameters)
%	- Load connected to a three-phase pi-line with in \dq coordinates, taken from CDC paper, with constant voltage input (const amplitude and frequency)
%	- Give system equations, system stability depends on load passivity
%		- Line ensures the system is not rigid, allows instability to occur if load not passive/system not stable
%	- Line parameters chosen ...
%	- Load parameters chosen as in table on both sides of the passivity boundary
%	- Simulated first with ZIP load, then with Exp load
%	- At first change, Zq, nq are reduced to the point of non-passivity at the voltage point of the load (t=0.5s) then Zp, np are increased to ensure passivity criteria are met (t = 0.6s)
%	- At second change, Zp, np are reduced to the point of non-passivity at the voltage point of the load (t=1.0s) then Zq, nq are used to passivate the system (t=1.1s)
%	- 
%	- 
% 3) Results (only factual)
%	- Stability, instability
% 4) Discussion of results (interpretation)
%	- 
%	- 
%	- Voltage and Frequency aymptotically stable despite only P controller
%	- Quickly returns to stable state after disturbance
%	- RMS voltage deviations small
%
% 1) Goal of Simulation: What is to be demonstrated?
In this section, we evaluate the results in \autoref{sec:Load_Passivity} by simulating a load node in \Matlab/\Simulink under various parameter configurations. 
We show that voltage stability (implying frequency stability under  \autoref{assumption:dq_synchrony}) is not guaranteed a priori as violating our sufficient conditions can lead to instability.
% 2) Description of the simulation setup: (Testsystem, events, inputsignals, parameters)
For this, we consider a voltage source at node $j$ connected via an electrical line to a load node $i$ as in \autoref{fig:load_circuit_model}. 
%By combining the load system PHS \eqref{eq:Modeling:loadphs} with the electrical line model in \cite[Eq. (13)]{Strehle19}, we obtain the system
%%
%\begin{equation}\label{eq:Simulation:model}
%	\begin{split}
%		\!\begin{bmatrix}
%			\cL[ij] \Iddot[ij] \\
%			\cL[ij] \Iqdot[ij] \\
%			0.5\cC[ij] \Vddot[j] \\
%			0.5\cC[ij] \Vqdot[j]
%		\end{bmatrix} \!\!=\!&
%		\begin{bmatrix}
%			-\cR[ij] & \freqRef \cL[ij] & -1 & 0 \\
%			-\freqRef \cL[ij] & -\cR[ij]  & 0 & -1 \\
%			1 & 0 & 0 & 0.5 \freqRef \cC[ij] \\
%			0 & 1 & - 0.5 \freqRef \cC[ij] & 0 \\
%		\end{bmatrix} \!\!\!\!\!\;
%		\begin{bmatrix}
%			\Id[ij] \\ \Iq[ij] \\ \Vd[j] \\ \Vq[j]
%		\end{bmatrix} \\ &+
%		\begin{bmatrix}
%			\Vd[i] \\ \Vq[i] \\ -\ILd(\Vdq[j]) \\ -\ILq(\Vdq[j])
%		\end{bmatrix} .
%	\end{split}
%\end{equation}
%
In contrast to the direct parallel connection of the load node $i$ with a voltage source, this creates a non-stiff system in which the load node can exhibit instability independent of the source supplying it. 
The voltage source is set to a constant amplitude of $\Vamp[j]=\SI{400}{\volt}$ for the \dq voltage vector $\Vdq[j]$. Any stabilizing or destabilizing behavior thus arises from within the system, i.e.\ the load behavior. The system is simulated with both a ZIP load and an exponential load, as per \eqref{eq:Modeling:current_ZIP} and \eqref{eq:Modeling:current_Exp}, respectively. The line and load parameters are described in \autoref{table:simulation_load_params}. The line parameters are derived from \cite{Strehle19} for a $\SI{1}{\kilo\meter}$ line. The loads parameters are changed during the course of the simulation to alternate between satisfying and violating the restrictions \eqref{eq:Load_Passivity:zip_restrictions_derivation} and \eqref{eq:Load_Passivity:exp_restrictions_derivation}, respectively.

% 3) Results (only factual)
% 	- results shown in figures, in each figure -> voltage limits 
The results of the simulation are shown in \autorefMulti{fig:Simulation:zip, fig:Simulation:exp}, respectively. In each figure, the voltage amplitude as per \eqref{eq:Load_Passivity:Load_currents:voltage_amp} is shown along with the changes made to the load parameters. Furthermore, sufficient voltage limits for strict passivity, which we calculated by solving \eqref{eq:Load_Passivity:zip_restrictions_derivation} and \eqref{eq:Load_Passivity:exp_restrictions_derivation} for the respective load parameters, are also indicated in the figures.
% 4) Discussion of results (interpretation)
Strict passivity and thus asymptotic voltage stability (implying frequency stability under \autoref{assumption:dq_synchrony}) is only guaranteed when the operating voltage amplitude $\Vamp[i]$ of the load node, induced by the input voltage $\Vdq[j]$, falls within this range. If these conditions are not met, the load could reverse the polarity of its dissipation, potentially resulting in instability. Note that due to the non-zero line impedance, changes to the ZIP load parameters, which result in differing current magnitudes, yield different operating points for $\Vamp[i]$ as in \autoref{fig:Simulation:zip}. However, since the exponents of exponential loads change the nature of the loads without the total power drawn at the nominal voltage, the variations in the points of operation are not seen in \autoref{fig:Simulation:exp}.
\begin{table}[!t]
	\centering
	\renewcommand{\arraystretch}{1.25}
	\caption{
		Simulation Parameter Values}
	\label{table:simulation_load_params}
	\begin{tabular}{@{\,\,}ll@{\!\quad}c@{\!\quad}c@{\!\quad}c@{\,\,}}
		\noalign{\hrule height 1.0pt}
		& Parameter & Value & \begin{tabular}{@{}c@{}} Change 1 \\ $(t=\SI{0.5}{\second})$ \\ $[t=\SI{0.6}{\second}]$\end{tabular} & \begin{tabular}{@{}c@{}} Change 2 \\ $(t=\SI{1.0}{\second})$ \\ $[t=\SI{1.1}{\second}]$\end{tabular} \\
		\hline
		\multirow{6}{*}{\begin{tabular}{@{}l@{}} ZIP \\ Load\end{tabular}} & $\lYp \, (\si{\per\ohm})$ & 0.15 & $\left(0.1\right)$ & $\left(0.2\right)$ \\
		& $\lYq \, (\si{\per\ohm})$ & 0.05 &  &  \\
		& $\lIp \, (\si{\ampere})$ & 2 &  &  \\
		& $\lIq \, (\si{\ampere})$ & 9 &  &  \\
		& $\lPp \, (\si{\kilo\watt})$ & 4.5 &  &  \\
		& $\lPq \, (\si{\kilo\VAR})$ & 19 & $\left[11\right]$ & $\left(24\right)$  \\
		\hline
		\multirow{4}{*}{\begin{tabular}{@{}l@{}} Exponential \\ Load\end{tabular}} & $\lActNom \, (\si{\kilo\watt})$ & 5.5 &  &  \\
		& $\lReacNom \, (\si{\kilo\VAR})$ & 3.7 &  &  \\
		& $\lNp$ & 1.7 & $\left(1.1\right)$ & $\left(1.3\right]$ \\
		& $\lNq$ & 0.7 & $\left[1.9\right]$ & $\left(0.45\right)$ \\
		\hline
		\multirow{3}{*}{Line} & $\cR[ij] \ (\si{\ohm})$ & 0.01273 &  & \\
		& $\cL[ij] \ (\si{\milli\henry})$ & 0.9337 &  & \\
		& $\cC[ij] \ (\si{\nano\farad})$ & 12.74 &  & \\
		%		\multirow{3}{*}{Line} & $\cR[ij] \ (\si{\ohm \per \km}) $  & 0.01273 & 0.3864 & \\
		%		& $\cL[ij] \ (\si{\milli\henry \per \km}) $ & 0.9337 & 4.1264 & \\
		%		& $\cC[ij] \ (\si{\nano\farad \per \km}) $ & 12.74  & 7.751 & \\
		\noalign{\hrule height 1.0pt}
	\end{tabular}
\end{table}
\begin{figure}[!t]
	\centering
	\resizebox{\columnwidth}{!}{%
		\tikzsetnextfilename{03_Img/simulation_zip}%
		\begin{tikzpicture}

\begin{axis}[%
width=4.521in,
height=1.493in,
at={(0.758in,2.554in)},
scale only axis,
unbounded coords=jump,
xmin=0,
xmax=1.5,
xlabel={Time in $\si{\second}$},
xtick={0,0.25,...,1.5},
ymin=360,
ymax=390,
ylabel={$\Vamp[i]$ in $\si{\volt}$},
ytick={360,370,...,390},
legend style={legend cell align=left, align=left}
/tikz/line join=bevel
]
\addplot [color=colorSim1,thick]
  table[row sep=crcr]{%
0.000740000000007512	390.007649060392\\
0.00111299999997527	378.946689944887\\
0.0014929999999822	376.397376487846\\
0.00187399999998661	375.81203648382\\
0.00225599999998849	375.677221114403\\
0.00263799999999037	375.646275922196\\
0.0030219999999872	375.639153963676\\
0.00341800000001058	375.637509380606\\
0.00390499999997473	375.637122110277\\
0.00521099999997432	375.637052484736\\
0.146825999999976	375.637052026786\\
0.5	375.637052026797\\
0.500000139615111	393\\
0.500078858946495	393\\
0.500079000000028	389.665591586025\\
0.500079030220945	393\\
0.500086130510113	393\\
0.500086467086817	357\\
0.500087421534545	357\\
0.500087687766893	393\\
0.500088678988504	393\\
0.500089000000003	373.03916383088\\
0.500089512374814	393\\
0.500090372729517	393\\
0.500090999999998	361.030331870169\\
0.500091519730177	393\\
0.50009231476389	393\\
0.500092698331514	357\\
0.500093165272062	357\\
0.500093375413371	393\\
0.500094535000017	393\\
0.500094714999989	357\\
0.500101583686444	357\\
0.500101583686444	357\\
0.500096494566549	357\\
0.500096862865462	393\\
0.50009685678333	393\\
0.500099284999976	357\\
0.500099465000005	393\\
0.50073589855208	393\\
0.500736000000018	387.841557113367\\
0.500736045992483	393\\
0.500739506827415	393\\
0.500740525305275	357\\
0.500740973513587	393\\
0.500740979714578	393\\
0.500740979714578	393\\
0.500742534999972	393\\
0.500742715000001	357\\
0.500743284999999	357\\
0.500743465000028	393\\
0.500762236535365	393\\
0.500762532070041	357\\
0.500763426147785	357\\
0.500763695293699	393\\
0.500763384749064	393\\
0.500763384749064	393\\
0.500765585928093	393\\
0.500765783062775	357\\
0.500763721071507	357\\
0.500767478470095	357\\
0.500767780661761	393\\
0.500767608377203	393\\
0.500769604012476	393\\
0.50076983819821	357\\
0.500770152603195	357\\
0.500770373475234	393\\
0.500773535000008	393\\
0.500774284999977	357\\
0.500774465000006	393\\
0.60018500000001	390.156489468667\\
0.600243999999975	362.105444794119\\
0.600256999999999	362.557647691215\\
0.600734999999986	381.363926493102\\
0.601050999999984	382.256632497578\\
0.601366999999982	382.371199496704\\
0.60168299999998	382.385833859849\\
0.602006000000017	382.387714247364\\
0.602449999999976	382.387960947461\\
0.606006999999977	382.387975414026\\
0.899999999999977	382.387975414024\\
0.900001535000001	393\\
0.900002285000028	357\\
0.900002465	393\\
0.900005832761792	393\\
0.900006000000019	371.511878434474\\
0.900006167238189	393\\
1.00100157251234	393\\
1.00100800000001	387.057631256875\\
1.00102199999998	389.971343304355\\
1.00102299999998	390.355006474319\\
1.002092	390.00236578389\\
1.00250599999998	376.506271236374\\
1.002928	372.623501229917\\
1.00335200000001	371.507020705374\\
1.00377700000001	371.1853417308\\
1.00420200000002	371.092798005571\\
1.00462700000003	371.066153921146\\
1.00505500000003	371.058454048259\\
1.005492	371.056233120893\\
1.00597900000002	371.055583197428\\
1.00679100000002	371.055396810856\\
1.01143999999999	371.055377783025\\
1.5	371.055377783016\\
};

\fill[colorSim1] (axis cs: 0.501,360) rectangle (0.600,390);
\fill[colorSim1] (axis cs: 0.901,360) rectangle (1.001,390);
\fill[greenLight, opacity=0.4] (axis cs: 0.0,360) rectangle (0.5,365);
\fill[greenLight, opacity=0.4] (axis cs: 0.6,360) rectangle (0.9,365);
\fill[greenLight, opacity=0.4] (axis cs: 1.0,360) rectangle (1.5,365);
\fill[redLight, opacity=0.5] (axis cs: 0.5,385) rectangle (0.6,390);
\fill[redLight, opacity=0.5] (axis cs: 0.9,385) rectangle (1.0,390);
\draw[dashed, line width=0.5,black!50!white](axis cs: 0.5,360) -- (axis cs: 0.5,390);
\draw[dashed, line width=0.5,black!50!white](axis cs: 0.6,360) -- (axis cs: 0.6,390);
\draw[dashed, line width=0.5,black!50!white](axis cs: 0.9,360) -- (axis cs: 0.9,390);
\draw[dashed, line width=0.5,black!50!white](axis cs: 1.0,360) -- (axis cs: 1.0,390);
%\fill[colorSim1] (axis cs: 0.50201,218) rectangle (0.66301,228);
%\fill[colorSim1] (axis cs: 1.0036358,218) rectangle (1.161,228);

\node[greenDark!60!black] at (axis cs: 0.25,362.3) {$\Vamp[i] > \SI{373}{\volt}$};
\node[greenDark!60!black] at (axis cs: 0.75,362.3) {$\Vamp[i] > \SI{363}{\volt}$};
\node[greenDark!60!black] at (axis cs: 1.25,362.3) {$\Vamp[i] > \SI{359}{\volt}$};

\node[redDark!60!black, anchor=east] at (axis cs: 0.5,387) {$\Vamp[i] > \SI{460}{\volt}$};
\node[redDark!60!black, anchor=west] at (axis cs: 1.0,387) {$\Vamp[i] > \SI{515}{\volt}$};

\end{axis}

\begin{axis}[%
width=4.521in,
height=1.493in,
at={(0.758in,0.481in)},
scale only axis,
xmin=0,
xmax=1.5,
xlabel={Time in $\si{\second}$},
xtick={0,0.25,...,1.5},
axis y line*=left, % '*' prevents arrow heads
ymin=0.08,
ymax=0.22,
ytick={0.1,0.15,...,0.2},
ylabel={Conductance $\lYp $ in $\si{\per\ohm}$},
legend style={legend cell align=left, align=left},
legend pos=south east
]
\addplot [color=colorSim1,thick]
  table[row sep=crcr]{%
0	0.15\\
0.5	0.15\\
0.5	0.1\\
1.0	0.1\\
1.0	0.2\\
1.5	0.2\\
};
\addlegendentry{$\lYp$}

\addplot [color=colorSim2,thick]
table[row sep=crcr]{%
0	7\\
};
\addlegendentry{$\lPq$}

\end{axis}

\begin{axis}[%
width=4.521in,
height=1.493in,
at={(0.758in,0.481in)},
scale only axis,
xmin=0,
xmax=1.5,
xlabel={Time (s)},
xtick={0,0.25,...,1.5},
axis y line*=right, % '*' prevents arrow heads
axis x line=none,
ymin=10,
ymax=25,
ytick={10,15,...,25},
ylabel={Power $\lPq $ in $\si{\kilo\watt}$},
legend style={legend cell align=left, align=left}
]

\addplot [color=colorSim2,thick]
  table[row sep=crcr]{%
0	19\\
0.6	19\\
0.6	11\\
0.9	11\\
0.9	24\\
1.5	24\\
};

\end{axis}
\end{tikzpicture}%
%
	}
	%	\scalebox{1.0}{\inputtikz{03_Img/simulation_zip}}
	\caption{Stability for various ZIP load configurations}
	\label{fig:Simulation:zip}
\end{figure}
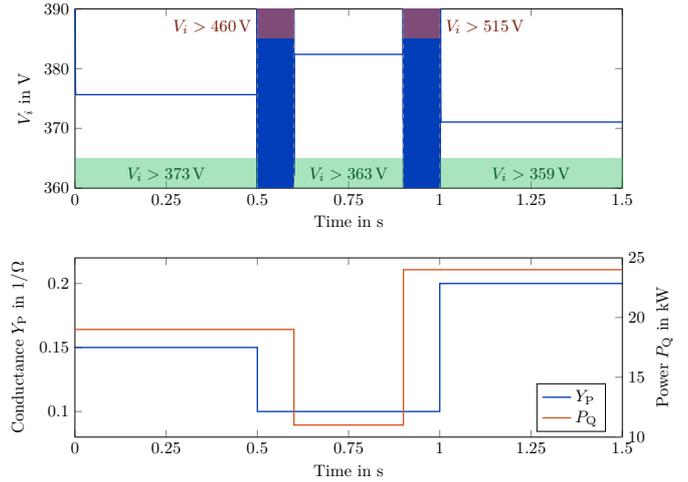
\begin{figure}[!t]
	\centering
	\resizebox{\columnwidth}{!}{%
		\tikzsetnextfilename{03_Img/simulation_exp}%
		\begin{tikzpicture}

\begin{axis}[%
width=4.521in,
height=1.493in,
at={(0.758in,2.554in)},
scale only axis,
unbounded coords=jump,
/tikz/line join=bevel,
xmin=0,
xmax=1.5,
xlabel={Time in $\si{\second}$},
xtick={0,0.25,...,1.5},
ymin=390,
ymax=410,
ylabel={$\Vamp[i]$ in $\si{\volt}$},
legend style={legend cell align=left, align=left}
]
\addplot [color=colorSim1,thick]
  table[row sep=crcr]{%
0	400\\
2.40000002804663e-06	388\\
5.99999998485146e-06	389.5\\
1.07000000184598e-05	412\\
0.00017400000001544	410.006812056247\\
0.000430999999991855	397.109196925451\\
0.000684999999975844	397.088133571991\\
0.00223499999998467	397.088099007218\\
0.5	397.088099007222\\
0.500001999999995	388.258761639943\\
0.500009999999975	389.81604996257\\
0.500014436790025	412\\
0.500021000000004	411.816849985333\\
0.50002576336999	388\\
0.500036036629979	388\\
0.500040836629978	412\\
0.500045	411.816849985333\\
0.500049763369987	388\\
0.500060541791242	388\\
0.500065000000006	410.29104377852\\
0.500067198715783	412\\
0.500072001284195	388\\
0.500076801284195	412\\
0.50007898715802	412\\
0.500082898715789	388\\
0.500095001284194	388\\
0.500100397431595	412\\
0.500103000000024	400.49357901192\\
0.500106000000017	411.493579011681\\
0.50061414999999	412\\
0.500616999999977	397.750000000251\\
0.500618999999972	399.750000000864\\
0.500621749999993	388\\
0.500626000000011	388.750000001102\\
0.500630849999993	412\\
0.500753499999973	412\\
0.500757000000021	395.000000000311\\
0.500760000000014	408.500000000421\\
0.50076439999998	388\\
0.500774299999989	388\\
0.500775999999973	392.999999999143\\
0.500777000000028	389.499999999875\\
0.500783285714306	388\\
0.50078400000001	390.499999999904\\
0.500784500000009	388\\
0.500806300000022	388\\
0.500809000000004	401.500000001885\\
0.500811999999996	391.000000001306\\
0.500812999999994	396.00000000145\\
0.500816199999974	388\\
0.500820999999974	411.000000002436\\
0.500825599999985	388\\
0.500889400000005	388\\
0.500894000000017	411.000000003547\\
0.501000926315783	412\\
0.501215000000002	405\\
0.625000999999997	405\\
0.625001073684189	412\\
0.626183000000026	410.499999999924\\
0.626187500000015	388\\
0.626260162576727	388\\
0.626264962576727	412\\
0.626331999999991	410.312984798181\\
0.626396	395.988683236182\\
0.626408000000026	396.104650971513\\
0.626505000000009	397.140341903264\\
0.626548000000014	397.120768392864\\
0.626614000000018	397.109968496776\\
0.626725000000022	397.110770111682\\
0.629185000000007	397.110749588547\\
0.899999999999977	397.110749588555\\
0.900001999999972	388.741682433226\\
0.900020051663489	388\\
0.900025000000028	410.741682433304\\
0.900029000000018	392.741682433341\\
0.900030000000015	393.241682433078\\
0.900031000000013	389.741682433255\\
0.900043451663521	388\\
0.900048000000027	410.179374498654\\
0.90039139999999	412\\
0.900392000000011	408.999999998493\\
0.90039274999998	412\\
0.900396100000023	412\\
0.900397999999996	403.499999998612\\
0.90039969999998	412\\
0.900420499999996	412\\
0.900423999999987	396.999999999258\\
0.900425999999982	403.500000000278\\
0.900426999999979	401.500000000775\\
0.900427999999977	403.000000001096\\
0.900431000000026	389.500000000985\\
0.900434000000018	389.000000001804\\
0.900439000000006	407.000000001211\\
0.900442800000008	388\\
0.900449400000014	388\\
0.900449999999978	391.000000001861\\
0.900450857142857	388\\
0.900453200000015	388\\
0.900459000000012	410.50000000381\\
0.900502000000017	411.500000004949\\
0.900503000000015	406.500000004805\\
0.900504000000012	410.500000004365\\
0.900819200000001	412\\
0.90082000000001	407.999999999574\\
0.900821500000006	412\\
0.900869999999998	410.50000000381\\
0.900874499999986	388\\
0.900918999999988	388\\
0.900923900000009	412\\
0.901000926315817	412\\
0.901214999999979	405\\
0.999001000000021	405\\
0.999001073684212	412\\
1.00037500000002	410.229291464182\\
1.000632	397.104928238045\\
1.00088599999998	397.082525988536\\
1.002229	397.082485085826\\
1.5	397.082485085832\\
};

\fill[colorSim1] (axis cs: 0.501,390) rectangle (0.625,410);
\fill[colorSim1] (axis cs: 0.901,390) rectangle (0.999,410);
\fill[greenLight, opacity=0.4] (axis cs: 0.0,390) rectangle (0.5,393.5);
\fill[greenLight, opacity=0.4] (axis cs: 0.6,390) rectangle (0.9,393.5);
\fill[greenLight, opacity=0.4] (axis cs: 1.0,390) rectangle (1.5,393.5);
\fill[redLight, opacity=0.5] (axis cs: 0.5,406.5) rectangle (0.6,410);
\fill[redLight, opacity=0.5] (axis cs: 0.9,406.5) rectangle (1.0,410);
\draw[dashed, line width=0.5,black!50!white](axis cs: 0.5,390) -- (axis cs: 0.5,410);
\draw[dashed, line width=0.5,black!50!white](axis cs: 0.6,390) -- (axis cs: 0.6,410);
\draw[dashed, line width=0.5,black!50!white](axis cs: 0.9,390) -- (axis cs: 0.9,410);
\draw[dashed, line width=0.5,black!50!white](axis cs: 1.0,390) -- (axis cs: 1.0,410);

\node[greenDark!60!black] at (axis cs: 0.25,391.75) {$\Vamp[i] > \SI{209}{\volt}$};
\node[greenDark!60!black] at (axis cs: 0.75,391.75) {$\Vamp[i] < \SI{6585}{\volt}$};
\node[greenDark!60!black] at (axis cs: 1.25,391.75) {$\Vamp[i] > \SI{377}{\volt}$};

\node[redDark!60!black, anchor=east] at (axis cs: 0.5,407.75) {$\Vamp[i] > \SI{899}{\volt}$};
\node[redDark!60!black, anchor=west] at (axis cs: 1.0,407.75) {$\Vamp[i] > \SI{863}{\volt}$};

\end{axis}

\begin{axis}[%
width=4.521in,
height=1.493in,
at={(0.758in,0.481in)},
scale only axis,
xmin=0,
xmax=1.5,
xlabel={Time in $\si{\second}$},
xtick={0,0.25,...,1.5},
ymin=0.25,
ymax=2.25,
ylabel={Power Index $n$},
ytick={0.5,1.0,...,2.0},
legend style={legend cell align=left, align=left}
]
\addplot [color=colorSim1,thick]
  table[row sep=crcr]{%
0	1.7\\
0.5	1.7\\
0.5	1.1\\
1.0	1.1\\
1.0	1.3\\
1.5	1.3\\
};
\addlegendentry{$\lNp$}

\addplot [color=colorSim2,thick]
  table[row sep=crcr]{%
0	0.7\\
0.6	0.7\\
0.6	1.9\\
0.9	1.9\\
0.9	0.45\\
1.5	0.45\\
};
\addlegendentry{$\lNq$}

\end{axis}
\end{tikzpicture}%
%
	}
	%	\scalebox{1.50}{\inputtikz{03_Img/simulation_exp}}
	%	\resizebox{\columnwidth}{!}{\includegraphics{03_Img/untitled.eps}}
	\caption{Stability for various exponential load configurations}
	\label{fig:Simulation:exp}
\end{figure}
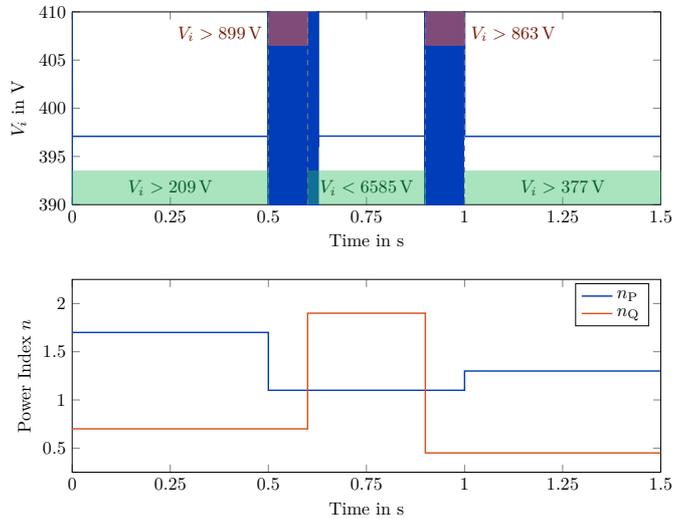
	\section{Conclusion}
In this work, we presented sufficient inequality conditions for the strict passivity of the prevalent nonlinear static ZIP and exponential AC load models. Together with our former results from \cite{Strehle19}, this allows us to infer asymptotic voltage and frequency stability of AC microgrids with arbitrary topologies in a plug-and-play manner via passivity arguments. 
Future research will include a more extensive simulative validation of our complete plug-and-play framework and investigate possibilities to broaden the feasible voltage operating area by providing necessary and sufficient inequality conditions.

\end{document}